\input epsf
\documentstyle[preprint,prl,aps]{revtex}
\begin{document}
\date{September 5th, 1997; Revised December 12th, 1997}
\draft
\title{NUCLEAR SHADOWING AND RHO PHOTOPRODUCTION}
\author{A.Pautz and G.Shaw}
\address{Theoretical Physics Group, Department of Physics and Astronomy, \\
The University of Manchester, M13 9PL, UK}
\maketitle
\begin{abstract}

Rho photoproduction on complex nuclei is re-examined using a 
generalised vector dominance model which succesfully predicts  
the  oberved nuclear
shadowing in real photoabsorption and deep inelastic scattering. This model 
 is shown to
give a good fit to rho photoproduction data on both nucleons and complex 
nuclei,  in which the  disagreement between the 
measured $\gamma-\rho$ coupling
and the $\gamma-\rho$ coupling required by the simple vector dominance model 
is eliminated.
The $\rho N$ total cross-sections required are similar to those predicted by 
the additive quark model;  and the magnitude of the correction to simple 
vector dominance is consistent with that inferred from the analysis of real
photoabsorption and deep inelastic scattering.

\end{abstract}


\section{Introduction}

In this paper, we address the apparent contradiction between the well-known
simple vector dominance(SVD) treatment of rho photoproduction; and the
very succesful generalized vector dominance(GVD) treatment of nuclear
shadowing in real photoabsorption and deep inelastic scattering.

In 1989, one of us \cite{Shaw89} pointed out that the observed qualitative 
features of 
nuclear shadowing in deep inelastic 
scattering \cite{Arneodo} are simply and naturally accounted for
in a generalized vector dominance(GVD) model \cite{Fraas75} which can be shown to
be dual to the
parton model \cite{Shaw89,Brodsky}. The crucial feature of this model, in 
the context of shadowing, is 
that the cross-sections  $\sigma_{VN}$
for scattering  a sequence of hadronic vector states 
 $V = \rho, \rho^{\prime}, \ldots$ from nucleons 
are required to be approximately
independent of their mass $m_V$. This then leads to 
approximate scaling behaviour for shadowing and a rapid decrease in
the effect as $x$ increases from zero, as observed in the 
data \cite{EMC}. Somewhat later, in 1993, the 
same model was  shown \cite{Shaw93a}\footnote{This paper was completed while 
the standard review \cite{Arneodo} of nuclear effects in structure functions 
was in press. It is therefore not covered in this 
review, despite the fact that it was published slightly earlier.}
to give an accurate quantitative account of both real photoabsorption on 
nuclei \cite{Caldwell} and the precise shadowing data that had by
then become  available for
virtual photons \cite{Shadowingdata}.

The above GVD model is consistent with the fundamental QCD picture of
strong interactions in appropriate kinematic regions\footnote{This is
discussed explicitly in recent papers \cite{Shaw93b} in which GVD models
are extended to incorporate the small-$x$ rise in the proton structure 
function associated with the ``hard Pomeron'', which is observed at 
large $Q^2$ at HERA. The resulting predictions for its behaviour at
low $Q^2$ have subsequently been confirmed by experiment \cite{F2data}.}. 
It has  implications for rho photoproduction, 
because it necessarily includes
substantial contributions from ``non-diagonal'' 
diffraction dissociation processes of the type $VN \rightarrow V^{\prime}N$
in addition to ``diagonal'' elastic processes of the type $VN \rightarrow VN$.
This feature is also implied by duality with the parton 
model \cite{Shaw89,Brodsky}. In addition,
without such processes, the cross-sections $\sigma_{VN}$ would be forced to 
decrease rapidly with mass $m_V$ to maintain approximate scaling in the 
nucleon structure functions; and  shadowing would  die 
away as $Q^2$ increases at fixed $x$, in contradiction to  approximate 
scaling for the nuclear structure functions.  These  
diffraction dissociation processes   also contribute to 
$rho$-photoproduction, as illustrated in 
Figure \ref{GVDdiagram}, giving significant corrections 
to the well-known predictions of the simple vector dominance(SVD) model.
This  result is potentially a serious problem, since rho photoproduction
on both nucleons and nuclei has long been regarded as the outstanding success
of the SVD model; and if the agreement between the SVD predictions and the
data were  really good, it would clearly undermine the above GVD approach.
However, as Donnachie and Landshoff \cite{DL95} have recently pointed out, 
the SVD predictions for the cross-sections on  nucleons are lie approximately 
16$\%$ above the measured values.

In this paper we shall investigate rho photoproduction on both nucleons and 
nuclei within the framework of the GVD model used to successfully account for
shadowing in real and virtual photoabsorption. The aims are to see whether 
model can resolve the discrepancy between the predictions of SVD and the
nucleon data, while retaining the successful predictions of SVD for
the $A$ dependence on nuclei; and if so, whether
sign and magnitude of the diffraction dissociation terms required by
the rho photoproduction data are consistent with those required by
the real photoabsorption and structure function data.

\section{Rho photoproduction on  nucleons}

We begin by reviewing rho photoproduction on nucleons 
$$ 
   \gamma + N \rightarrow \rho + N  \hspace{2cm} (N = n, \, p)
$$
-a topic in which there is renewed interest because  of recent 
data from HERA {\cite{ZEUS,H1}.
At high energies, the process
is usually described in terms of the SVD
model of Figure \ref{GVDdiagram}, leading  to the well-known 
relation\footnote{For an review of vector meson 
photoproduction and other diffractive photoprocesses in the context of SVD, 
see Leith \cite{Leith78}.}
\begin{equation}
 f_{\gamma  \rho }(s, t=0) = \frac{e}{f_{\rho}} f_{\rho \rho }(s, t=0)
\hspace{1.5cm}  (SVD)
\label{i0} 
\end{equation}
where $f_{ab}$ is the scattering amplitude for $a + N \rightarrow b + N$
and $f_{\rho}$ is the $\gamma- \rho$ coupling. Using the optical theorem,
this gives
\begin{equation}
\frac{d \, \sigma}{d \, t} (s , t=0) = \alpha \frac{4 \, \pi}{f_{\rho}^2}
\sigma_{\rho N}^2 \, [1 + \eta^2] \hspace{1.5cm}  (SVD)
\label{i1}
\end{equation}
for the forward differential cross-section, where
 $\sigma_{\rho N}$ is the total cross-section for $\rho \, N$
scattering and $\eta$ is the ratio of the real to imaginary part of the 
forward  $\rho \, N$ scattering amplitude. The value of 
$\sigma_{\rho N}$ is usually taken from the additive quark model
prediction
\begin{equation}
\sigma_{\rho \, N} = \frac{1}{2} \left[ \sigma_{\pi^+ \, N} +
\sigma_{\pi^- \, N} \right] \; ;
\label{i2}
\end{equation}
and $\eta$ is estimated from Regge pole ideas. 

The resulting SVD  
prediction (\ref{i1}, \ref{i2})
was long ago  compared with experimental photoproduction data on protons
available at energies below 20 GeV to yield a value \cite{Leith78}
\begin{equation}
\frac{f_{\rho}^2}{ 4 \, \pi} = 2.44 \pm 0.12 \; .
\label{i3}
\end{equation}
for the coupling constant. At the time,  this simple picture was 
consistent with the corresponding SVD 
analysis of vector meson photoproduction on nuclei 
\cite{Alvensleben,McClellan} 
and with the then not very precisely known value of 
the $\gamma- \rho$ coupling
obtained directly from the measured decay 
width $\Gamma (\rho \rightarrow e^+e^-)$.
Since then events have moved on and  it is important to check 
whether this consistency still obtains. The precision of
the measured decay width $\Gamma (\rho \rightarrow e^+e^-)$
has greatly increased \cite{PDG}, and now gives 
\begin{equation}
\frac{f_{\rho}^2}{ 4 \, \pi} = 2.01 \pm 0.10 \; ,
\label{i4}
\end{equation}
which is not in good agreement with the phenomenological value (\ref{i3}).
This problem is confirmed by a recent comparison \cite{DL95}
incorporating  higher energy photoproduction data \cite{ZEUS,Aston}, which finds that  
the SVD prediction
(\ref{i1}, \ref{i2}) obtained using the directly measured coupling (\ref{i4})
lies on average about 16 \% above the measured data.

The discrepancy between the experimental data and the SVD approximation 
 presumably arises from contributions which are
neglected in this approximation. There are two obvious possibilities.  The 
first is  finite width effects associated with 
the $\rho^0 \rightarrow \pi^+ \; \pi^-$ decay channel, as shown in 
Figure \ref{twopi}. Detailed calculations 
show that this is very unlikely, since the correction arising from the
modification to the rho propagator is largely cancelled by corrections
arising from the pion scattering terms \cite{Bauer76}. 

The second possibility
is to attribute the discrepancy to neglected contributions from
higher mass vector states, leading to the generalized vector dominance(GVD) 
model shown in Figure \ref{GVDdiagram}. 
In principle many states could contribute, but in practice only the lightest 
states are expected to be important, since  the diffraction dissociation 
amplitudes $f(\rho^{\prime} N \rightarrow \rho N)$, and to a lesser degree 
the couplings $(e / f_{\rho^{\prime}})$, are expected to 
decrease rapidly with increasing mass $m_{\rho^{\prime}}$ \cite{Donnachie78}. 
Hence it is reasonable to  approximate  these contributions with that of a
single effective $\rho^{\prime}$ state, when the SVD prediction (\ref{i0}) is  
replaced by
\begin{equation}
f_{\gamma \rho} =\frac{e}{f_{\rho}}f_{\rho \rho}  +
 \frac{e}{f_{\rho^{\prime}}}f_{\rho^{\prime} \rho}   \hspace{1.5cm}
(GVD) \, .
\label{gammarho}
\end{equation}
This form  also follows directly from the generalised vector dominance model 
\cite{Fraas75} used to successfully predict the observed shadowing effects 
in both real 
photoabsorption and deep inelastic scattering on nuclei \cite{Shaw93a}.
We shall not discuss this  model
further, but just use it to specify the 
properties of the effective $\rho^{\prime}$, which are
\begin{equation} 
m_{\rho^{\prime}}^{2}=3m_{\rho}^{2} \ \ \ \ \ \ \ \ \ \ \ f_{\rho^{\prime}}=
m_{\rho^{\prime}} f_{\rho} / m_{\rho}  \label{vectormeson}
\end{equation}
together with
\begin{equation} 
f_{\rho^{\prime}  \rho^{\prime} }= f_{\rho \rho }  \hspace{2cm}
f_{\rho^{\prime} \rho }  = f_{\rho \rho^{\prime}}  = 
- \epsilon f_{ \rho  \rho }
\label{vectormeson2}
\end{equation}
for the forward scattering amplitudes, where 
\begin{equation}
 \epsilon \approx 0.2 \; .
\label{epsilon2}
\end{equation}
Here, we shall retain (\ref{gammarho} - \ref{vectormeson2}) but
treat $\epsilon$ as a free parameter to be determine by
the $\rho$-photoproduction data. The 
forward cross-cross-section is then  given by 
\begin{equation} \label{GVD}
\frac{d \, \sigma}{d \, t} (s , t=0) = \alpha \frac{4 \, \pi}{f_{\rho}^2}
\sigma_{\rho N}^2 \, [1 + \eta^2]
\left[1-\epsilon \frac{m_{\rho}}{m_{\rho^{\prime}}}\right]^{2} \hspace{1cm}
(GVD). 
\end{equation}
For fixed $\sigma_{\rho N}$ and $\eta$, the results of SVD with the effective 
coupling (\ref{i3}) are 
reproduced by GVD with the physical coupling (\ref{i4}) and 
$ \epsilon=0.160\pm 0.055$, which is in good agreement with the value
(\ref{epsilon2}) quoted above.

We thus have two accounts of the data on $\rho$ photoproduction on protons:
either the SVD model in which the $\gamma-\rho$ coupling is 
adjused to fit the data; or the GVD model in which  this coupling
 is fixed at its measured value  and 
and the   parameter $\epsilon$ is  adjusted to fit the data.
The latter is more compatible with the ideas used to understand shadowing in
nucleon structure functions, but its 
predictions for $\rho$ photoproduction on nuclei have never been examined. 
It is our
aim to remedy this omission and discuss the implications of the results.

\section{Rho photoproduction on nuclei}

In the generalised vector dominance model,
an incident photon  may convert into a 
a whole sequence of  isovector, vector
mesons $V = \rho, \rho^{\prime}, \rho^{\prime \prime}, \ldots$
while traversing a nucleus. According to Glauber theory \cite{Glauber}
\cite{Grammer}, 
this possibility is dealt with by introducing appropriate optical potentials\ 
$U_{\gamma V}$, $U_{\gamma \gamma}$, $U_{V V}$, $U_{V V^{\prime}}$, which 
characterise the 
features of the scatterer, where $V,V^{\prime}$ are arbitrary members of the
vector meson sequence. 
The resulting wave equation then reads
\begin{equation}
\left[ \left( \begin{array}{cc}
        \bigtriangledown^{2}+k_{\gamma}^{2} & {\bf 0 }\\
        {\bf 0}   & \cal V                                
        \end{array} \right)-
        \left(
        \begin{array}{cc}
        U_{\gamma \gamma}  & \mbox{$\boldmath U_{\gamma}^{T}$}   \\
        \mbox{$\boldmath U_{\gamma}$} & \cal U
        \end{array}                    
        \right) \right]
        \left( \begin{array}{c}
        \Psi_{\gamma} \\ {\bf \Psi}
        \end{array} \right)=0
\end{equation}
where 
\begin{eqnarray}
\left(\cal V\right)_{V V^{\prime}} & = & \delta_{V V^{\prime}}(\bigtriangledown^{2}+k^{2}_{V}) \\ 
\left(\cal U\right)_{V V^{\prime}} & = & U_{V V^{\prime}}\nonumber \\
\left({\bf \Psi}\right)_{V}   &   =   &  \psi_{V} \nonumber   \\
\left(\mbox{$\boldmath U_{\gamma}$}\right)_{V}  &  =  & U_{\gamma V}\ .\nonumber
\end{eqnarray}

In this paper we restrict ourselves to two hadronic channels 
$\rho, \rho^{\prime}$ and use the eikonal approximation
\begin{eqnarray} 
\Psi_{\gamma}=\Phi_{\gamma}e^{ik_{\gamma}z}\ \ \ \ & & 
\Psi_{\rho,\rho^{\prime}}=\Phi_{\rho,\rho^{\prime}}e^{ik_{\rho,\rho^{\prime}}z}
\label{eikonal}
\end{eqnarray}
where the reduced wave functions $\Phi_{\gamma}$, $\Phi_{\rho}$, $\Phi_{\rho^{\prime}}$ are  assumed to vary 
slowly enough for all  second derivatives to be discarded.
Furthermore, since we are only interested in  the photoproduction amplitude, 
it is sufficient to work  to order $\cal O(\sqrt{\alpha})$ only. Hence, when
the  photon wavefunction occurs multiplied by factors $U_{\gamma\rho}$ or 
$U_{\gamma\gamma}$,
which are already of this or higher order, we can replace it by the 
incident photon wave $\Psi_{\gamma}= e^{ik_{\gamma}z}$.
With these approximations, the reduced wavefunctions of the
vector mesons    satisfy the
coupled differential equations
\begin{equation}\label{de1}
\ \ \! \frac{d}{dz}\Phi_{\rho}({\bf b},z)=-\frac{i}{2k_{\rho}}\left(U_{\gamma\rho}e^{iq_{\|_{\gamma\rho}}z}+
U_{\rho\rho}\Phi_{\rho}({\bf b},z)+U_{\rho\rho^{\prime}}\Phi_{\rho^{\prime}}({\bf b},z)e^{iq_{\|_{\rho\rho^{\prime}}}z}\right)
\end{equation}
\begin{equation} 
\frac{d}{dz}\Phi_{\rho^{\prime}}({\bf b},z)=-\frac{i}{2k_{\rho^{\prime}}}\left(U_{\gamma\rho^{\prime}}e^{iq_{\|_{\gamma\rho^{\prime}}}z}+
U_{\rho^{\prime}\rho^{\prime}}\Phi_{\rho^{\prime}}({\bf b},z)+U_{\rho^{\prime}\rho}\Phi_{\rho}({\bf b},z)e^{iq_{\|_{\rho^{\prime}\rho}}z}\right)
\label{de2}
\end{equation}
where the space coordinates are parameterized through the z-coordinate in 
the direction of the incident photon wave and the impact 
parameter {\bf b} in the plane perpendicular
to the z-axis. The expressions 
$q_{\|_{ij}}=k_{i}-k_{j}$ denote the minimal 
longitudinal momentum transfer between particle i and j, where i and j are 
members of the set $\gamma,\ \rho,\ \rho^{\prime}$.

\subsection{Optical potentials and nuclear densities.}

The optical potentials are given by \cite{Glauber}
\begin{eqnarray}
U_{ij}=-4\pi f_{ij }\ n({\bf b},z)\ \ \ \ \ &  & U_{ij}=U_{ji}
\label{pots}
\end{eqnarray}
where the forward scattering amplitudes on nucleons $f_{ij}$
are given by eqs.(\ref{gammarho} - \ref{vectormeson2})
together with the GVD analogue of (\ref{gammarho})
for the $\rho^{\prime}$:
\begin{equation}
f_{\gamma \rho^{\prime}}
=\frac{e}{f_{\rho}}f_{\rho \rho^{\prime}} +
 \frac{e}{f_{\rho^{\prime}}}f_{\rho^{\prime} \rho^{\prime}}   \, .
\label{gammarhop}
\end{equation}
Two different models \cite{Grammer} are employed for
 the nuclear density  $n({\bf b},z)$, 
depending on the size of the nucleus. 
\begin{itemize}
\item{$A>16$}
\newline
For heavy nuclei,\ $ A>16$, we use a Fermi gas-like mass distribution of the form
\begin{equation}
n({\bf r})=n_{W}\left(1+\exp\left(\frac{r-R_{W}}{c}\right)\right)^{-1}\ .
\label{Woods}
\end{equation}
where the \lq skin-thickness\rq  $ \; c=0.545 $ fm. The Woods-Saxon 
radius $R_W$
is given as the solution of the equation
\begin{equation}
A\ f(R_{W})=n_{W} \equiv A_{\mbox{\scriptsize Pb}}\ f(R_{\mbox{\scriptsize Pb}})\label{findwoods}
\end{equation}
corresponding to a fixed central density $n_W$, where 
\begin{equation}
f(\rho)=\left(\frac{3}{4\pi \rho^{3}}\right)\left(1+\pi^{2} \frac{c^{2}}{\rho^{2}}\right)^{-1}
\end{equation}
and $R_{Pb} = 6.626$ fm.
\item{$A<16$}
\newline
For light nuclei, the Woods-Saxon formula is not  a good description. Instead we
use a shell model (harmonic oscillator) density, given by
\begin{equation}
n(r)=n_{s}\left(1+\delta\left(\frac{r}{R_{S}}\right)^{2}\right)
\exp\left(-\left(\frac{r}{R_{S}}\right)^{2}\right)
\end{equation}
where 
\begin{eqnarray}
\delta=\frac{A-4}{6}\ \ \ \ \ \ \   &  
\ \ \ \ \ \ \  n_{s}=\frac{2A}{2+3\delta}\left(\sqrt{\pi} R_{S}\right)^{-3}\label{shell}
\end{eqnarray}
and the shell model radius 
 $R_S =  0.708 \,\mbox{A}^{\frac{1}{3}}$ fm.
\end{itemize}
Finally, to take two-body correlations into account, 
we modify the nuclear density functions by the replacement \cite{spital2}
\begin{equation}
n({\bf r})\ \longrightarrow\ n({\bf r})\left(1+\frac{1}{2}\ l_{c}\ \sigma_{\rho N}
\ n({\bf r})\ \left(\frac{n({\bf r}=0)}{n({\bf r})}\right)^{-\frac{1}{3}}\right)
\end{equation}
where  the two body correlation length  $l_{c}$
=0.3 fm.

\subsection{Evaluation of the cross-sections}

To evaluate the cross-section for $\rho$ photoproduction on heavy nuclei,
we need to solve  (\ref{de1}, \ref{de2}) with the initial conditions
\begin{equation}
\Phi_{\rho}({\bf b},z  = - \infty) = 
\Phi_{\rho^{\prime}}({\bf b},z= - \infty) =0
\label{initial}
\end{equation}
corresponding to incident photons. The forward differential cross section   
is then given by
\begin{equation}
\frac{d\sigma}{dt}(t=0)\ =\ \frac{\pi}{k_{\rho}^{2}}\ |F_{\gamma\rho}(0)|^{2}\ .\label{dsigma}
\end{equation}
where the forward scattering amplitude $F_{\gamma\rho}$ is related to the 
\lq\lq profile function''
\begin{equation}
\Gamma_{\rho}({\bf b})=-\lim_{z \rightarrow \infty}\Phi_{\rho}({\bf b},z)\ .  \label{profile}
\end{equation} 
by the standard result   \cite{Glauber}  
\begin{equation}
F_{\gamma\rho}(0)=i\, k_{\rho}\int_{0}^{\infty}db\ b\ \Gamma(b) \; .
\end{equation}

From   (\ref{de1}, \ref{de2}), it is clear  that the reduced 
wavefunctions only vary with $z$ at fixed impact parameter ${\bf b}$
in regions where optical potentials  (\ref{pots}) are non-zero.
Since the nuclear densities fall off very rapidly beyond the nuclear radii,
we neglect such variations for $|z| > 4 R_W$, measured from the centre of the
nucleus. The initial conditions (\ref{initial}) are imposed 
at $ z = - 4 R_W$ and (\ref{de1}, \ref{de2}) are integrated up to
 $ z = + 4 R_W$ at fixed impact
parameter ${\bf b}$ using a  fourth-order Runge-Kutta algorithm with an
 adaptive stepsize control. The profile function is then evaluated at
 $ z = + 4 R_W$ rather than infinity and used to compute  cross-sections,
which are compared with experiment in the concluding section. 
Typical results for the reduced wavefunction 
 $\Phi_{\rho}$ and the profile function $\Gamma (b)$ are shown in 
Fig. \ref{fig:wavy} and Fig. \ref{fig:impact1} respectively, verifying the
assumed disappearance of structure by $ 4 R_W$.

\subsection{A simple approximation}

Before comparing with experiment, we  comment briefly   on an 
approximation that gives a useful check on our numerical solutions.
In the SVD model, the eikonal equation (\ref{de1}) 
with $f_{\rho \rho^{\prime}} = 0$ can be integrated explicitly  and this 
result
is easily generalized to the  coupled channel GVD 
case (\ref{de1}, \ref{de2}) in the 
high energy limit when
\begin{equation}
{\rm e}^{i q_{\|_{\rho \rho^{\prime}}} z} = 1 \; .
\label{approx1}
\end{equation}
This approximation has been used to analyse the $\rho$ photoproduction 
data by Kroker \cite{Kroker} giving, for example, a value 
 $\sigma_{\rho N} = 29.0 \pm 1.2$ mb corresponding to an incident photon energy 
of 6.1 GeV. The corresponding prediction for the nuclear cross-sections is 
shown  in Figure \ref{compare}, where it is compared both to the data and 
to our
\lq\lq exact'' predictions for the same parameter values. In fact 
$$
{\rm e}^{i q_{\|_{\rho \rho^{\prime}}} z} = i \hspace{1cm} {\rm for}
\hspace{1cm} E_{\gamma} = 6.1 \, GeV , \; \; \; z \approx 3 \, f \; .
$$
 Since  $z = 3$ f is well within a  large nucleus 
for small impact parameters $b \approx 0 $,
it is  clear  that (\ref{approx1}) is quantitatively unreliable at the 
energies 
where data currently exists. We shall not consider it further, except to note
that  our 
numerical solutions reproduce both the standard SVD predictions and
Kroker's approximate results \cite{Kroker} 
when we impose  $f_{\rho \rho^{\prime}} = 0$ or (\ref{approx1}) respectively.

\section{Results and Conclusions}

In this section we present the results of an optical model analysis of the
 experimental data on complex nuclei using the GVD model described. 
At each energy the cross-sections depend on four parameters: 
$$
  \frac{f_{\rho}^2}{4 \pi}  \hspace{0.5cm}   \sigma_{\rho N}  \hspace{0.5cm}
 \epsilon  \hspace{0.5cm}  \eta  \; .
$$
Our strategy is to fix the coupling $f_{\rho}$ at its measured value (\ref{i4})
and the phase ratio  $\eta$ at the values implied by the Regge pole 
parameterization 
\begin{equation}
f_{\rho \rho}(s,t=0)=g_{P}\frac{-1-e^{-\pi \alpha_{P}}}{\sin(\pi \alpha_{P})}s^{\alpha_{P}}+
        g_{R}\frac{-1-e^{-\pi \alpha_{R}}}{\sin(\pi \alpha_{R})}s^{\alpha_{R}}
\label{phase}
\end{equation}
where the subscripts P and R denote the pomeron and Regge contributions 
respectively and the constants $\alpha_P, \; \alpha_R, \; g_P$ and $g_R$ are 
fixed by assuming  the additive 
quark model relation $f_{\rho N}=f_{\pi N}$ together with the
Donnachie-Landshoff fit \cite{Donnachie93} to  $\pi N$  scattering data. 
The remaining 
parameters $\sigma_{\rho N}$ and $ \epsilon$
 are then determined by fitting to the $\rho$-photoproduction 
data\footnote{ The contrasting 
treatment of the cross-section  $\sigma_{\rho N}$ and phase $\eta$  is 
justified because the results are relatively
insensitive to small changes in the phase, which is about $\eta = 0.2$ at
the energies of the data used, but very sensitive to $\sigma_{\rho N}$.}. 
First, however, we summarise the  earlier SVD results and 
 the data available. 
 
We are only interested in data at high enough   energies for  
the phase parameter $\eta$ to be small and reasonably well estimated by
conventional Regge pole ideas.
The most precise data on complex nuclei at such energies was obtained by a 
DESY-MIT group \cite{Alvensleben} 
at a photon energy of 6.6 GeV and by a Cornell group\cite{McClellan}) at
6.1, 6.5 and 8.8 GeV. The latter group also presented results  on
protons and
deuterium, giving  measurements of the single nucleon cross-section  from
the same experiment. They then carried out an SVD analysis of their own 
 nucleon and  nuclear  cross-sections, assuming input phases $\eta$
which are somewhat larger than those given by (\ref{phase}), 
and treating both the $\gamma-\rho$ coupling and the 
 $\rho N$ total cross-section as parameters to be fitted to the data.    
Their results are summarised in Table I. This analysis was repeated by
Kroker \cite{Kroker} with phases close to those assumed here, and
combining both the Cornell and DESY-MIT data at 6.5/6.6 GeV.  
The resulting   parameter values are 
given in Table II, and  the corresponding prediction  
is compared with the data at 6.1 GeV in Figure \ref{compare}.
Since both analyses used exactly the same data at 6.1 and 
8.8 GeV, a comparison of Tables I and II at these energies explicitly confirms
the insensitivity of the results  to small changes in the 
input phases. 

Here, we  repeat this anaysis using the GVD model described above, using
the experimental $\gamma-\rho$ coupling value  (\ref{i4}). Since this 
involves non-trivial numerical computation,
we simplify the determination of the two free parameters $\epsilon$ and
$\sigma_{\rho N}$ by
requiring that the GVD predictions  exactly reproduce the succesful SVD
results on single nucleons; and then determine the remaining parameter 
by a fit to the nuclear data. Satisfactory fits are obtained in this way 
at all three energies 6.1,6.5 and 8.8 GeV as shown in  
Figures \ref{6100}, \ref{6500} and \ref{8800}. The corresponding 
parameter values are shown in Table III. As can be seen, the  values 
obtained for the off-diagonal parameter $\epsilon$ at the three energies are 
consistent with each other and with
the value (\ref{epsilon2}) required by the successful treatment of nuclear
shadowing in real photoabsorption and deep inelastic scattering \cite{Shaw93a}.
The central values for $\sigma_{\rho N}$ are slightly larger than  those 
obtained in the SVD model fits(cf. Table II), but are consistent within the
quoted uncertainties arising from the  errors on the experimental data.
They are also slightly larger than, but in qualitative agreement with, 
the crude prediction (\ref{i2}) of the aditive quark model, which gives
$\sigma_{\rho N} \approx 27,27,26$ mb at $E_{\gamma} = 6.1,6.5,8.8$ GeV
respectively.

In short, the model is in good agreement with the data 
on $\rho$-photoproduction on both nucleons and nuclei, 
with a $\gamma-\rho$
coupling consistent with that measured in electron-positron annihihation.
The form and magnitude of the correction to simple vector dominance,
characterized by the parameter $\epsilon$, is consistent with that required
by the succesful description of shadowing in both real photoabsorption and 
deep inelastic scattering.       


\newpage
\begin{center}
\bf{TABLE CAPTIONS}
\end{center}

{\bf Table I} 
Parameters of the SVD fits of the Cornell group  [16] to their own 
data, 
together with the input values of the phase ratio $\eta$.

{\bf Table II}
Parameters of the SVD fits of Kroker [24] to the Cornell data [16] 
at 6.1 and 8.8 GeV, and the combined Cornell and DESY-MIT data [15] at 6.5 GeV,
together with the input values of the phase ratio $\eta$.

{\bf Table III}
Parameters of our optical model GVD fits to the Cornell data [16] 
at 6.1 and 8.8 GeV, and the combined Cornell and DESY-MIT data [15] at 6.5 GeV.
The phase ratio $\eta$ is given by (29) and
the $\gamma-\rho$ coupling is fixed at its  measured value (5).

\newpage
\begin{center}
\bf{FIGURE CAPTIONS}
\end{center}

{\bf Figure 1}
The SVD diagram for $\rho^{0}$ photoproduction({\em left}), together
with the GVD corrections to it arising from
diffraction dissociation terms({\em right}).

{\bf Figure 2}
Finite width corrections to SVD associated with the
$2\pi$- channel. The first diagram is a propagator correction, the second a
$2\pi$- scattering contribution.

{\bf Figure 3}
The reduced wave function $\Phi_{\rho}$ calculated at zero
impact parameter for lead nuclei at 6.1 GeV. The parameters are those of our
final fit(see Table III below.)

{\bf Figure 4}
The profile function $\Gamma_{\rho}$ calculated for lead nuclei at 
6.1 GeV. The parameters are those of our
final fit(see Table III below).

{\bf Figure 5}
Comparison between different theoretical predictions and the 
complex nuclei data for
$E_{\gamma}$=6.1 GeV: 
the SVD-prediction assuming the parameter values of Table II 
(dashed dotted line);
the GVD  solution in the approximation (28) using the parameter
values given by Kroker [24] (dashed line);
and the full GVD prediction using the same parameters (solid line).

{\bf Figure 6}
The optical moddel GVD fit to the complex nuclei data at 6.1 GeV, 
corresponding to
the parameters of Table III.(The deuteron data point  shown  is  
not included in the  fit for obvious reasons.)

{\bf Figure 7}
The optical model GVD fit to the complex nuclei data at 6.5 GeV, 
corresponding to
the parameters of Table III.(The deuteron data point  shown is  
not included in the  fit for obvious reasons.)

{\bf Figure 8}
The optical model GVD fit to the complex nuclei data at 8.8 GeV, 
corresponding to
the parameters of Table III.(The deuteron data point  shown is  
 not included in the  fit for obvious reasons.)

\newpage

\begin{table}
\begin{center}
\begin{tabular}[t]{|c|c|c|c|}\hline
 $E_{\gamma}$ [MeV] & $\alpha_{\eta}$ &$\sigma_{\rho N}\ \ $[mb]
 & $ f_{\rho}^{2} / 4\pi $ \\ \hline
6.1 & -0.27 & 27.5$\pm$1.1 & 2.48$\pm$0.16   \\
6.5 & -0.27 & 27.9$\pm$1.3 & 2.60$\pm$0.20  \\
8.8 & -0.24 & 25.9$\pm$1.0 & 2.52$\pm$0.16   \\ \hline
\end{tabular}
\end{center}
\caption{Parameters of the SVD fits of the Cornell group  [16] to their own 
data, 
together with the input values of the phase ratio $\eta$.} 
\label{SVDparameters1}
\end{table}

\begin{table}
\begin{center}
\begin{tabular}[t]{|c|c|c|c|}\hline
 $E_{\gamma}$ [MeV] & $\eta$ &$\sigma_{\rho N}\ \ $[mb]
 & $ f_{\rho}^{2} / 4\pi$ \\ \hline
6.1 & -0.24 & 27.5$\pm$1.1 & 2.44$\pm$0.16   \\
6.5 & -0.23 & 26.7$\pm$2.0 & 2.28$\pm$0.10  \\
8.8 & -0.20 & 26.2$\pm$1.0 & 2.52$\pm$0.16   \\ \hline
\end{tabular}
\end{center}
\caption{Parameters of the SVD fits of Kroker [24] to the Cornell data [16] 
at 6.1 and 8.8 GeV, and the combined Cornell and DESY-MIT data [15] at 6.5 GeV,
together with the input values of the phase ratio $\eta$.}
\label{SVDparameters2}
\end{table}

\begin{table}
\begin{center}
\begin{tabular}[b]{|c|c|c|c|c|}\hline
 $E_{\gamma}$ [MeV] & $\eta$ &$\sigma_{\rho N}\ \ $[mb] & 
$f_{\rho}^{2} / 4\pi $ & $\epsilon$ \\ \hline
6.1 & -0.26 & 28.1$\pm$1.1 & 2.01 & 0.19$\pm$0.050  \\
6.5 & -0.24 & 28.4$\pm$2.0 & 2.01 & 0.21$\pm$0.035  \\
8.8 & -0.20 & 27.9$\pm$1.0 & 2.01 & 0.28$\pm$0.046   \\ \hline
\end{tabular}
\end{center}
\caption{Parameters of our optical model GVD fits to the Cornell data [16] 
at 6.1 and 8.8 GeV, and the combined Cornell and DESY-MIT data [15] at 6.5 GeV.
The phase ratio $\eta$ is given by (29) and
the $\gamma-\rho$ coupling is fixed at its  measured value (5).
}
\label{GVDparameters3}
\end{table}

\begin{figure}[t]
\begin{center}
\mbox{
\epsfxsize=14.5cm
\epsfbox{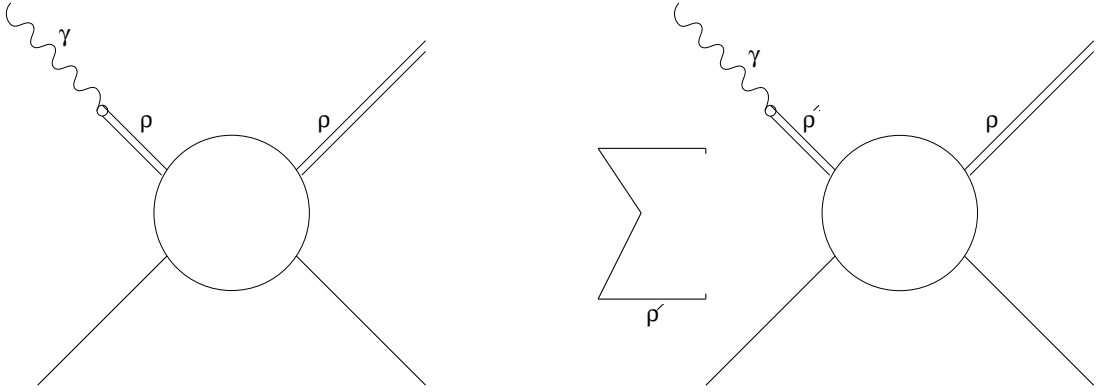}
}
\end{center}
\caption{The SVD diagram for $\rho^{0}$ photoproduction({\em left}), together
with the GVD corrections to it arising from
diffraction dissociation terms({\em right}).}
\label{GVDdiagram}
\vspace{0.8cm}
\end{figure}

\begin{figure}[t]
\begin{center}
\mbox{
\epsfxsize=13cm
\epsfbox{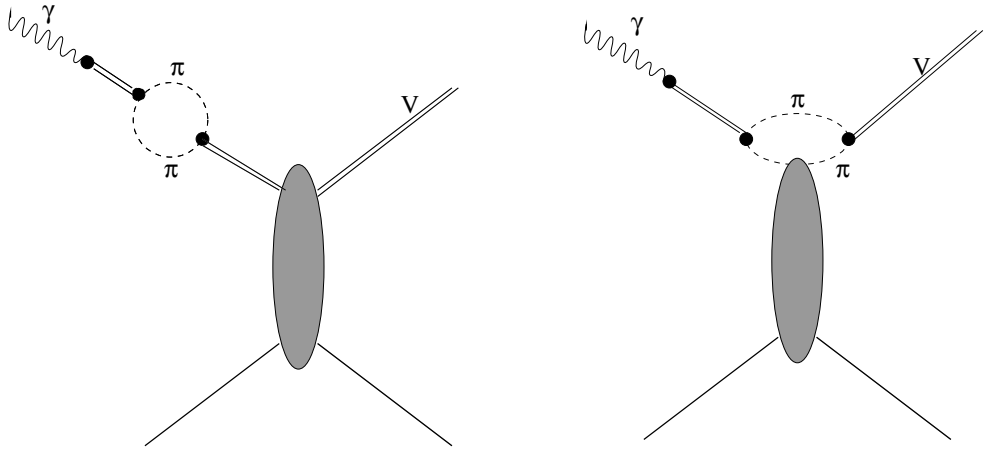}
}
\end{center}
\caption{Finite width corrections to SVD associated with the
$2\pi$- channel. The first diagram is a propagator correction, the second a
$2\pi$- scattering contribution.}
\vspace{0.8cm}
\label{twopi}
\end{figure}

\begin{figure}[t] 
\begin{center}
\mbox{
\epsfxsize=14.8cm
\epsfbox{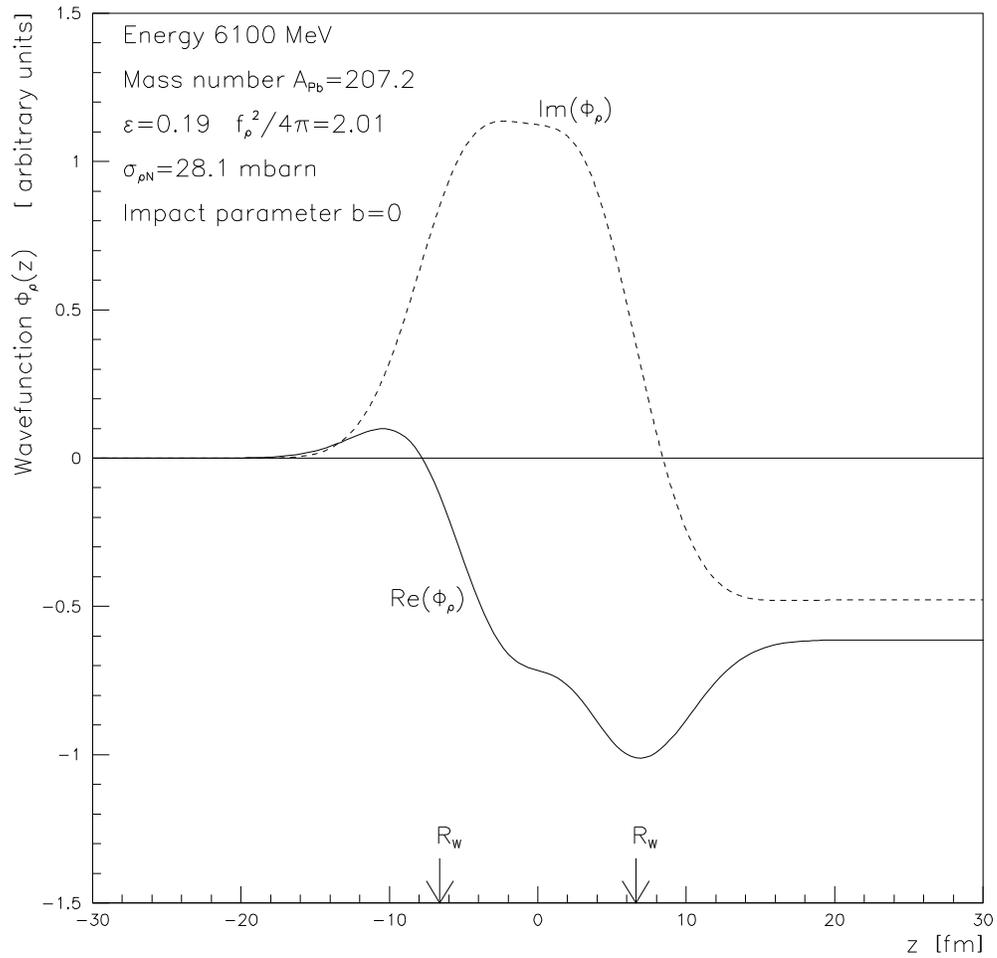}
}
\end{center}
\caption{The reduced wave function $\Phi_{\rho}$ calculated at zero
impact parameter for lead nuclei at 6.1 GeV. The parameters are those of our
final fit(see Table III below.)}
\label{fig:wavy}
\end{figure}

\begin{figure}[h] 
\begin{center}
\mbox{
\epsfxsize=14.8cm
\epsfbox{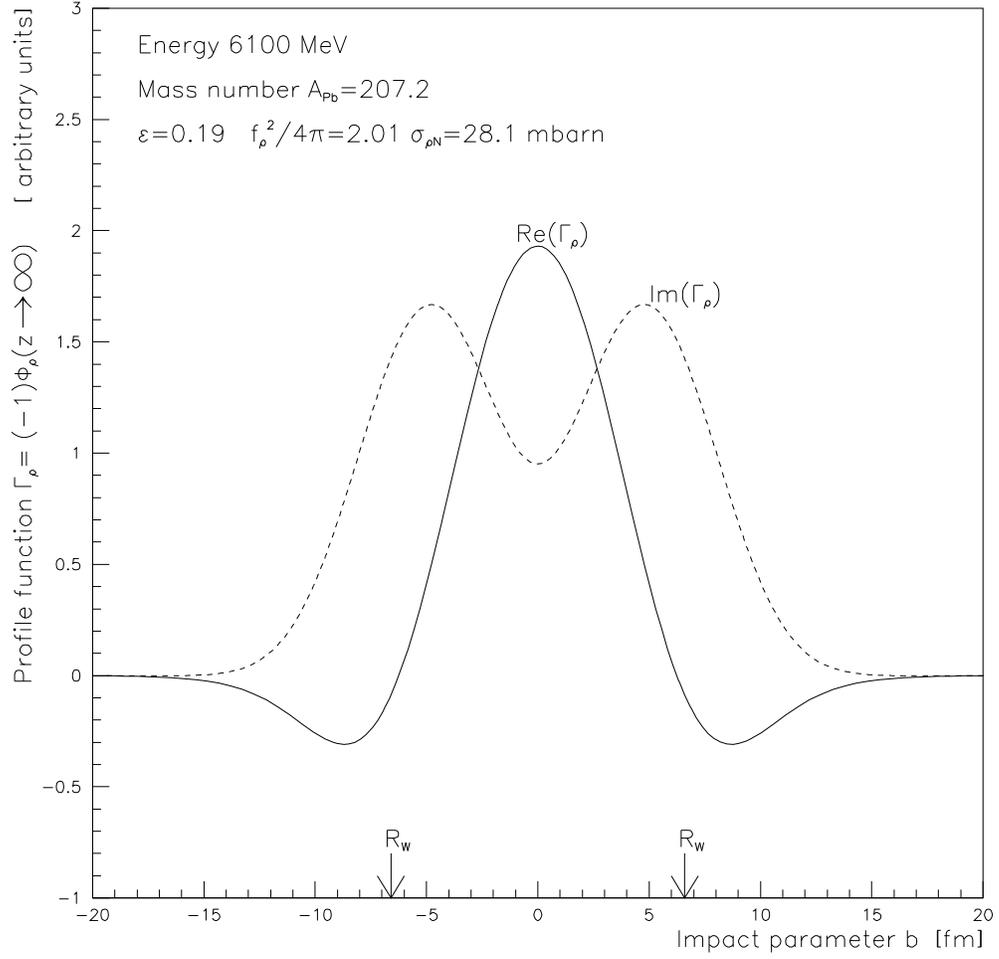}
}
\end{center}
\caption{The profile function $\Gamma_{\rho}$ calculated for lead nuclei at 
6.1 GeV. The parameters are those of our
final fit(see Table III below).}
\label{fig:impact1}
\end{figure}

\begin{figure}[t] 
\begin{center}
\mbox{
\epsfxsize=14.8cm
\epsfbox{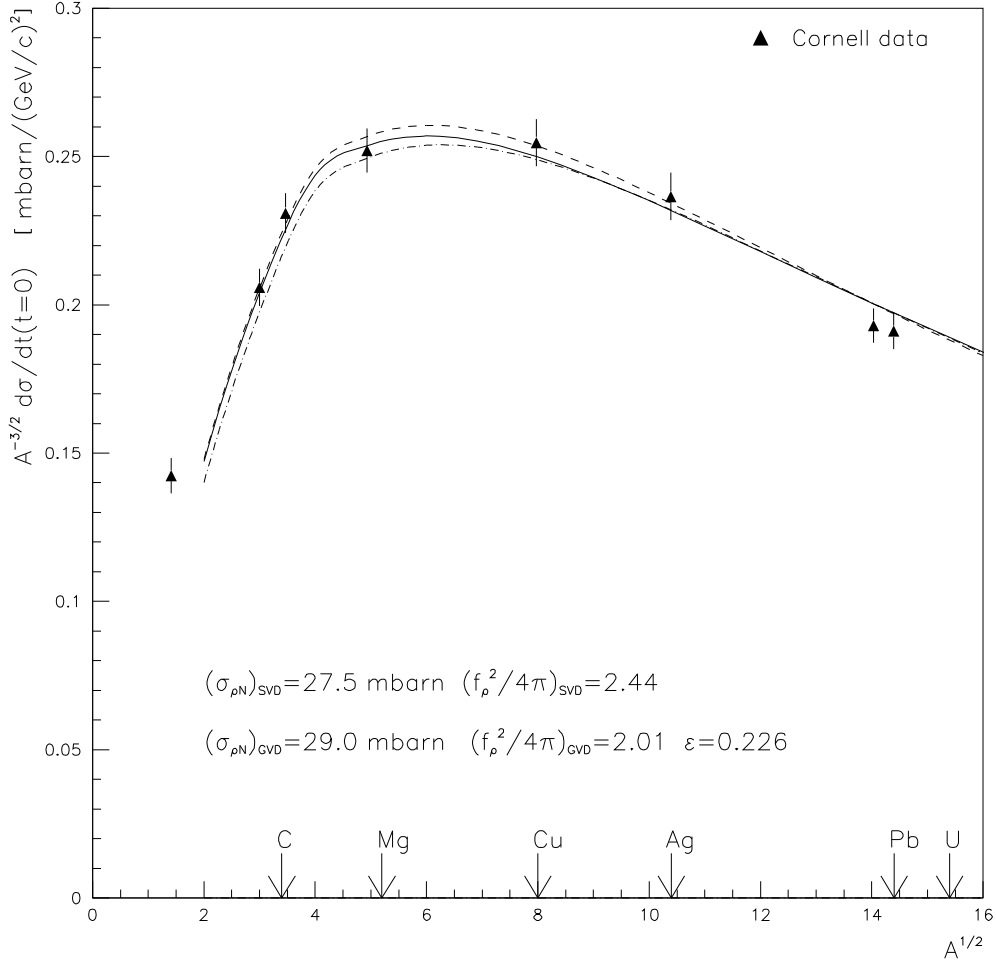}
}
\end{center}
\caption{Comparison between different theoretical predictions and the 
complex nuclei data for
$E_{\gamma}$=6.1 GeV: 
the SVD-prediction assuming the parameter values of Table II 
(dashed dotted line);
the GVD  solution in the approximation (28) using the parameter
values given by Kroker [24] (dashed line);
and the full GVD prediction using the same parameters (solid line)
.}
\label{compare}
\vspace{0.8cm}
\end{figure}

\begin{figure}[t]
\begin{center}
\mbox{
\epsfxsize=14.8cm
\epsfbox{lanlfigs/6100bestfit.prnt}
}
\end{center}
\caption{The optical moddel GVD fit to the complex nuclei data at 6.1 GeV, 
corresponding to
the parameters of Table III.(The deuteron data point  shown  is  
not included in the  fit for obvious reasons.) }
\label{6100}
\vspace{0.8cm}
\end{figure}

\begin{figure}[t]
\begin{center}
\mbox{
\epsfxsize=14.8cm
\epsfbox{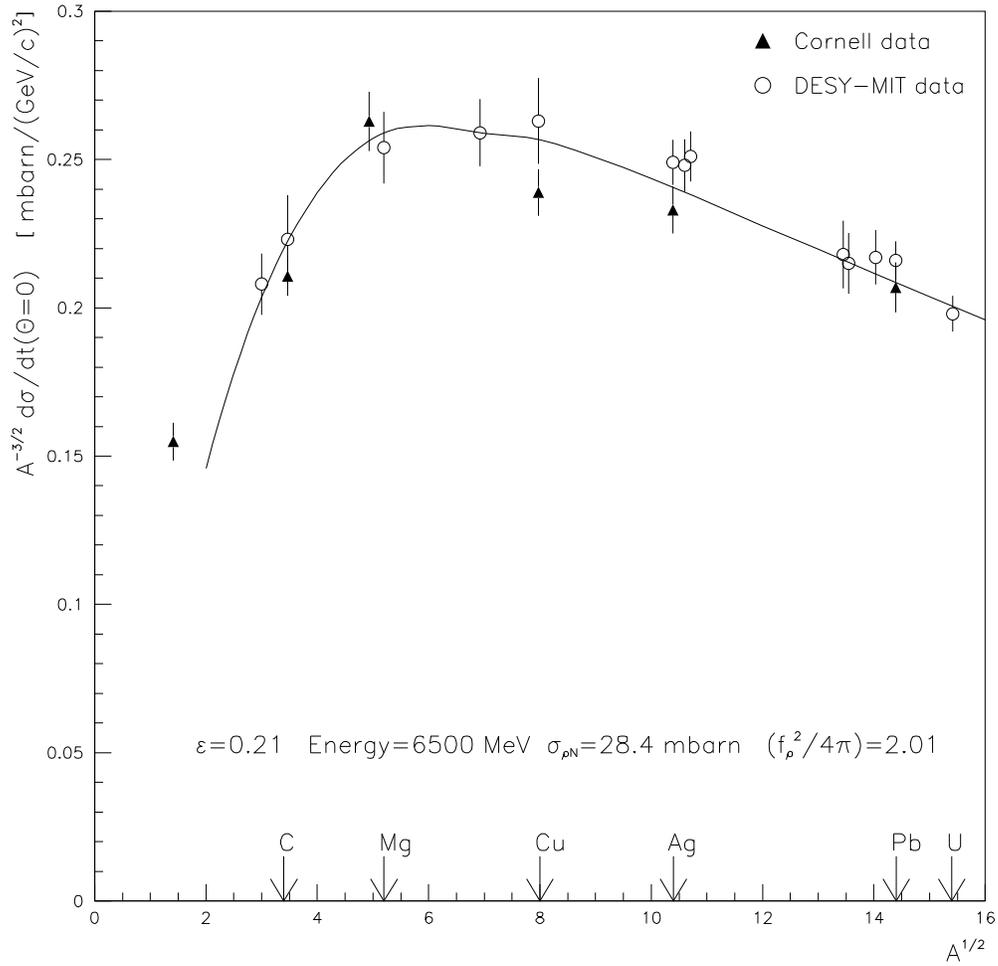}
}
\end{center}
\caption{The optical model GVD fit to the complex nuclei data at 6.5 GeV, 
corresponding to
the parameters of Table III.(The deuteron data point  shown is  
not included in the  fit for obvious reasons.) }
\label{6500}
\vspace{0.8cm}
\end{figure}

\begin{figure}[t]
\begin{center}
\mbox{
\epsfxsize=14.8cm
\epsfbox{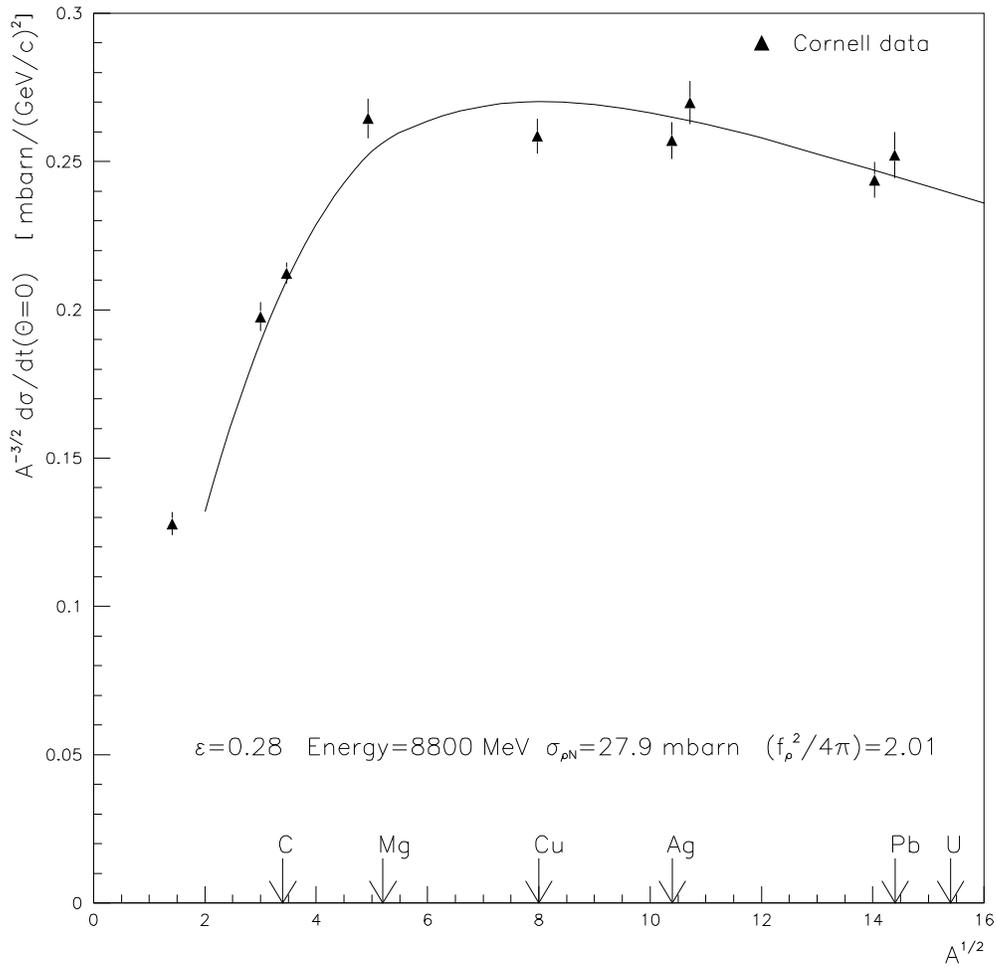}
}
\end{center}
\caption{The optical model GVD fit to the complex nuclei data at 8.8 GeV, 
corresponding to
the parameters of Table III.(The deuteron data point  shown is  
 not included in the  fit for obvious reasons.) }
\label{8800}
\vspace{0.8cm}
\end{figure}

\end{document}